\documentclass[]{emulateapj}
\usepackage{natbib}

\shorttitle{Multiple populations in young massive star clusters}
\shortauthors{Peacock et al.}

\newcommand{\hst}{{\it HST} }
\newcommand{\Ang}{${\rm \AA}$}
\newcommand{\Ha}{H$_{\alpha}$ }
\newcommand{\Hb}{H$_{\beta}$ }
\newcommand{\oii}{[O~{\sc ii}] }
\newcommand{\oiil}{[O~{\sc ii}]$\lambda$~3727 }
\newcommand{\oiill}{[O~{\sc ii}]$\lambda\lambda$~3726,3729 }
\newcommand{\oiii}{[O~{\sc iii}] }
\newcommand{\oiiil}{[O~{\sc iii}]$\lambda$~5007 }
\newcommand{\oiiill}{[O~{\sc iii}]$\lambda\lambda$~4959,5007 }

\begin{document}

\title{Signatures of multiple stellar populations in unresolved extragalactic globular/ young massive star clusters}

\author{Mark B. Peacock$^{1}$, Stephen E. Zepf$^{1}$ and Thomas Finzell$^{1}$}
\affil{$^{1}$Department of Physics and Astronomy, Michigan State University, East Lansing, MI 48824, USA}
\email{MBP: mpeacock@msu.edu}

\begin{abstract}
\label{sec:abstract}

We present an investigation of potential signatures of the formation of multiple stellar populations in recently formed extragalactic star clusters. All of the Galactic globular clusters for which good samples of individual stellar abundances are available show evidence for multiple populations. This appears to require that multiple episodes of star formation and light element enrichment are the norm in the history of a globular cluster. We show that there are detectable observational signatures of multiple formation events in the unresolved spectra of massive, young extragalactic star clusters. We present the results of a pilot program to search for one of the cleanest signatures that we identify -- the combined presence of emission lines from a very recently formed population and absorption lines from a somewhat older population. A possible example of such a system is identified in the Antennae galaxies. This source's spectrum shows evidence of two stellar populations with ages of 8~Myr and 80~Myr. Further investigation shows that these populations are in fact physically separated, but only by a projected distance of 59~pc. We show that the clusters are consistent with being bound and discuss the possibility that their coalescence could result in a single globular cluster hosting multiple stellar populations. While not the prototypical system proposed by most theories of the formation of multiple populations in clusters, the detection of this system in a small sample is both encouraging and interesting. Our investigation suggests that expanded surveys of massive young star clusters should detect more clusters with such signatures. 

\end{abstract}


\section{Introduction}
\label{sec:intro}

Traditionally, globular clusters have been considered as simple stellar populations (SSPs) -- produced from a single short burst of star formation, resulting in a population of stars with similar ages and abundances. This view follows naturally from the narrow color-magnitude diagrams observed in Galactic globular clusters. The well-defined main-sequence turnoffs limit any age spread, while the narrowness of features such as the red giant branch limit abundance spreads, at least in iron \citep[e.g.][and references therein]{Ashman98}. Further evidence that the spread in iron-peak abundances is small comes from detailed spectroscopic studies of individual stars in Galactic globular clusters \citep[e.g. ${\rm \Delta\lbrack Fe/H \rbrack}<$0.05~dex for stars in 19 Galactic clusters,][]{Carretta09b}. The few globular clusters that differ from this picture, by having clear spreads in iron-peak abundances among their stars (e.g.\ $\omega$ Cen, M22, M54, Terzan5), are typically the most massive globular clusters and only represent a few percent of the overall globular cluster population \citep[see e.g.\ the review's of][and references therein]{Gratton04,Gratton12}.

However, it is now clear that multiple populations are the norm in the Galactic globular clusters. A key component in this revolution in understanding globular clusters is the significant star-to-star variation observed in light elements such as C, N, O, Na and Mg \citep[see e.g.][and the many references therein]{Carretta10, Gratton12}. Initial hints of variations in the abundances of light elements came from the spectra of small samples of giant stars observed in clusters \citep[e.g.][]{Kraft94}. Subsequently, spectroscopy of much larger samples of globular cluster stars, including main sequence stars, has found the same variations in light element abundances. The presence of these variations among unevolved stars within the same globular cluster clearly indicates that these variations were present in the gas at the time the stars formed in the clusters. 

Notably, the variations in light elements such as C, N, O, Na and Mg have been seen within all of the many globular clusters now observed in sufficient detail \citep[e.g.][and the many references therein]{Carretta10, Gratton12}. Thus, these variations represent a fundamental aspect of globular cluster formation. Moreover, the variations are not random but have clear patterns. Prominent among these is the well-known Na - O anticorrelation, such that stars with an overabundance of Na have an underabundance of O. A similar anti-correlation is seen in Al and Mg. Such anti-correlations are a signature of H-burning at high temperatures \citep[e.g.][]{Langer93}, and thus place strong constraints on models for the origin of the material out of which the second population forms. It is also notable that stars with unusual patterns of light elements make up a substantial fraction of all stars within a globular cluster. Specifically, about half of the stars in globular clusters have abundance ratios similar to those seen in halo stars, while the other half or so have different light element ratios, with for example a range of enhanced N and reduced O \citep[e.g.][]{Carretta09a, Carretta10, Gratton12}. Furthermore, there is a fairly smooth distribution of enhanced Na and correspondingly depleted O among the stars with unusual abundances -- from just slightly higher [Na/O] than typical halo stars to an order of magnitude (or more) enhancement.

A number of other signatures of multiple stellar populations have been observed in the Galactic globular clusters. One example is the multiple or broadened main sequences that are being observed in an increasing number of globular clusters \citep[see e.g.][and references therein, for a discussion of clusters currently known to exhibit such signatures]{Piotto12}. These multiple sequences are often explained via a second generation of stars that have enhanced He abundance. Theories aimed at understanding the ubiquitous light element variations observed in globular clusters generally predict such an enhancement. Spectroscopy of stars in the different sequences appears to confirm that they are associated with the observed abundance variations \citep[e.g.][]{Bragaglia10}. Variations in horizontal branch morphology have also been tied to multiple stellar populations, particularly through variation in He \citep[e.g.][and references therein]{Gratton13}. Additional evidence for multiple populations in star clusters has been presented based on \hst observations of Large Magellanic Cloud (LMC) clusters \citep{Mackey07, Milone09, Goudfrooij11, Rubele13}. These observations demonstrated a broad or split main sequence turn-off in these intermediate age clusters, possibly suggesting an age spread of a few hundred Myr. 

The presence of multiple populations in clusters has challenged our understanding of their formation. The variation in light element abundances within every globular cluster studied to date indicates that the multiple population phenomena is a fundamental property of globular clusters. The nature of the variations requires the second generation to be formed out of material enriched by hot H-burning. As discussed in the following section, the leading explanation appears to be the formation of a second generation of stars out of ejecta from the cluster's asymptotic giant branch (AGB) stars mixed with some pristine material \citep[see e.g.][]{DAntona07, DErcole08, Conroy10, Conroy12}. 

The goal of this paper is to help test ideas for the origin of the multiple populations by identifying and studying the signature of multiple populations in recently formed massive star clusters. Many such clusters have been discovered in the local universe, however most are too distant to be resolved into individual stars. Therefore, our emphasis is to develop signatures of multiple stellar populations in young, unresolved star clusters. In Section \ref{sec:signatures}, we use stellar population synthesis models to investigate the signatures of multiple stellar populations on the integrated photometry (\ref{sec:signatures_color}) and spectroscopy (\ref{sec:signatures_spectra}) of a star cluster. Section \ref{sec:stats} discusses the probability of detecting evidence of multiple stellar populations in observations of young, massive star clusters based on the main signature we present. Finally, we present in Section \ref{sec:obs}, a pilot program to search for such signatures from star clusters in the Antennae galaxies.

\section{Signatures of multiple populations in unresolved star clusters}
\label{sec:signatures}

The stellar populations of star clusters beyond our Local Group can not generally be resolved. Therefore, in order to study multiple stellar populations in extragalactic globular clusters, we need to identify their effect on the integrated emission from a cluster. In this section we consider the differences between the emission from a single stellar population and multiple stellar populations and discus their potential detectability. In the following discussion we take the simplest case of a multiple stellar population -- that of a cluster with two distinct bursts of star formation. The age difference between the bursts of star formation in multiple population clusters is observationally quite poorly constrained. Observations of star clusters in the Magellanic Clouds show spreads in the main sequence turnoffs that, if due solely to an age difference, imply an age spread of several hundred Myr \citep[e.g.][]{Goudfrooij11,Rubele10}. However, other effects such as rotation or He abundances may also act to broaden the main sequence. Therefore, this constraint represents an upper limit on the age difference between populations in these clusters. 

Theoretical considerations imply certain constraints on the age difference between the two populations. The spectroscopic evidence from the Galactic globular clusters requires that the second population must be formed from gas enriched by H-burning at high temperatures. Currently, the leading contender for producing the polluted second generation gas is intermediate mass ($\sim$4-8$M_{\odot}$) AGB stars. In this scenario, hot H-burning occurs at the base of the convective envelope and is convected to the surface. This gas is then released into the intercluster medium via winds or with the ejection of the stars envelope. To produce the correct abundances in the second generation gas, this scenario requires AGB stars with masses in the range $\sim$4-8$M_{\odot}$ and implies that the second generation should be $\sim$40-200~Myr older than the first \citep[e.g.][]{DAntona07}. Fast rotating massive stars (20-120$M_{\odot}$, FRMS) have also been proposed as potential polluters. Here, hot H-burning occurs in the core of the star, its products are driven to the surface via rotational mixing and expelled via an equatorial wind into the intercluster medium. However, this scenario, which invokes massive stars and hence short timescales, has difficulties in producing new stars which are not enriched in iron peak elements (from supernovae) and which have the correct light element pattern \citep[e.g.][]{Renzini08, Conroy11}. Moreover, such models can not account for the possible age spreads suggested by the young LMC clusters. 

\begin{figure*}
 \centering
 \includegraphics[height=170mm,angle=270]{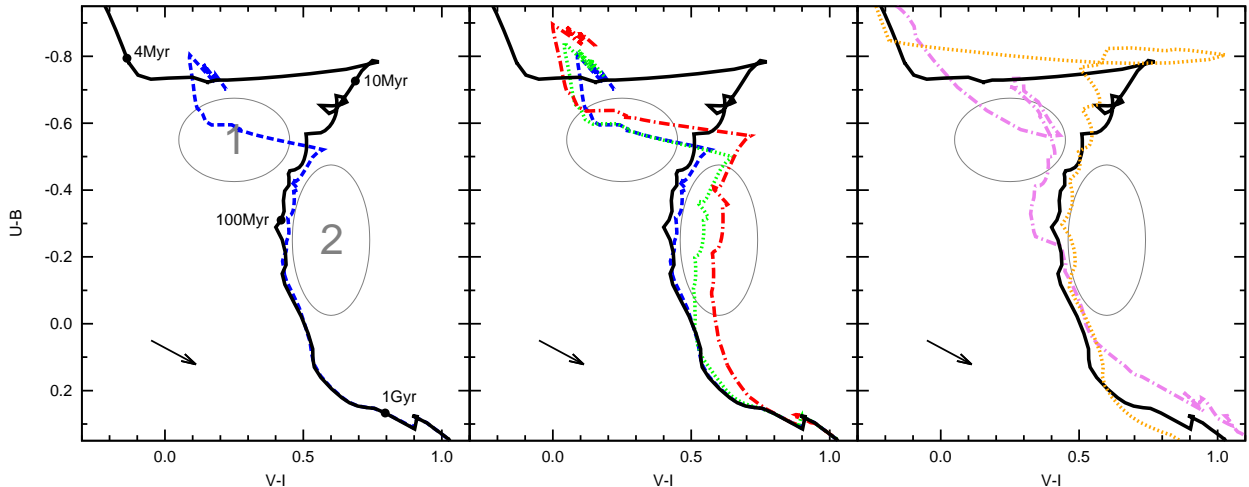} 
 \caption{The evolution of a star cluster's U-B vs. V-I colors from the \citet{Bruzual03} stellar population models. The black lines show a SSP with solar metallicity (aging from top left to bottom right). The blue dashed line shows a similar population, but with an additional burst of star formation, 100~Myr after the first. The second generation of stars is assumed to provide 10$\%$ of the cluster's total mass. The middle panel shows the same models but with different age spreads between the two bursts of star formation: 100~Myr (blue dashed line); 250~Myr (green dotted line); and 500~Myr (red dot-dashed line). The right panel shows SSP models with [Fe/H]=-0.33~dex (purple dot-dashed line), 0.00~dex (black solid line) and 0.56~dex (orange dotted line). Highlighted are the two regions in this color space where the multiple stellar populations have significantly different colors from SSPs. The arrow represents a reddening of E(B-V)=0.1~mag.} 
 \label{fig:bc03_colors} 
\end{figure*}

Spectroscopic observations of individual stars in the Galactic globular clusters have found that the second generation of stars comprise a large fraction of the total stellar population in these clusters, around 50$\%$ \citep[e.g.][]{Carretta09a}. However, it should be noted that at earlier times the fraction of second generation stars is likely to have been much lower. Indeed, leading theories for the formation of multiple populations in clusters often requires this \citep[e.g.][]{DErcole08}. This is because the second generation gas is polluted by the first generation of stars, which makes it very hard to produce a second generation which is $\gtrsim$10$\%$ of the total cluster mass \citep[e.g.][]{Bekki06}. However, if the first generation of stars is preferentially lost, then this ratio can increase to that observed as the cluster evolves. Such a scenario may occur if the second generation of stars is more centrally concentrated than the first. Simulations have demonstrated that this process can produce a cluster with similar fractions of first to second generation stars after a Hubble time \citep[e.g.][]{DErcole08, Bekki11}. 

\subsection{Broadband optical colors: U-B vs. V-I} 
\label{sec:signatures_color}

In Figure \ref{fig:bc03_colors} we plot the broadband U-B and V-I colors of a star cluster, produced using the \citet{Bruzual03} stellar population synthesis code. In the left panel, we consider the tracks traced by a solar metallicity stellar population as a function of its age. The black line shows a SSP formed from a single burst of star formation. The cluster colors trace a path from top left to bottom right as the stellar population evolves. The biggest departure from this general trend occurs after a few Myr, when the cluster reddens rapidly in V-I due to the evolution of massive stars to supergiants. These models are found to be in good agreement with the observed U-B and V-I colors of young massive star clusters in both the Antennae \citep{Whitmore10} and M83 \citep{Chandar10}. 

The blue dashed line in Figure \ref{fig:bc03_colors} shows the color of the same cluster, but with an additional burst of star formation 100~Myr after the first. For these illustrative purposes we consider this second population to be formed with $10\%$ of the total cluster mass, this is in the range predicted at early stages in the cluster's evolution (as discussed above). The second panel shows the same cluster, but with different times between the bursts of star formation (100, 250 and 500~Myr). It can be seen that by ages $\gtrsim$1~Gyr, the two populations in these clusters sum to give very similar U-B and V-I colors to a cluster with a SSP. Therefore, such observations at these ages would be unable to distinguish a multiple from a single stellar population. 

Some differences can be observed in this color space. For a few Myr after the second burst of star formation the cluster colors become bluer and fill region 1 of the color magnitude diagram. Such clusters would lie off the tracks of a SSP and this shift would be detectable with accurate photometry. Young, massive star clusters in the Antennae galaxies are found to lie in this region \citep[e.g.][]{Whitmore10}. However, several factors could result in SSP clusters also having colors in this region. Firstly, reddening can cause SSP clusters of a few Myr to be shifted to these colors. Also, as can be seen from the right panel of Figure \ref{fig:bc03_colors}, SSP clusters with sub-solar metallicity can have similar colors. A broader multiwavelength approach may be able to account for these factors. However, there are also variations between the \citet{Bruzual03} models and other commonly used stellar population models, such as the Starburst99 model \citep{Leitherer99} and that of \citet{Maraston05}. While other SSP models predict the period of rapid reddening of the a cluster's V-I color, the tracks vary slightly and not all show SSPs avoiding region 1. The other period where differences are observed between single and multiple population clusters is around a few hundred Myr (region 2 in Figure \ref{fig:bc03_colors}). Here, massive giant stars from the second generation result in redder V-I colors than would be produced from a SSP cluster, but have little effect on the cluster's U-B color. A different metallicity SSP cluster could not explain such a shift, although reddening of a SSP cluster potentially could. Such a shift could be reliably measured, but only if the age difference between the two stellar populations was large (even at 500~Myr, the shift is only V-I=0.2~mag). Interestingly, no young massive star clusters (with ${\rm M_{V}}$$<$-10) in the Antennae or M83 are found with such colors \citep{Whitmore10,Chandar10}. A dearth of clusters in this region may argue against multiple stellar population clusters with large age spreads of more than a few hundred Myr (in agreement with current formation theories). 

\begin{figure}
 \centering
 \includegraphics[height=120mm,angle=0]{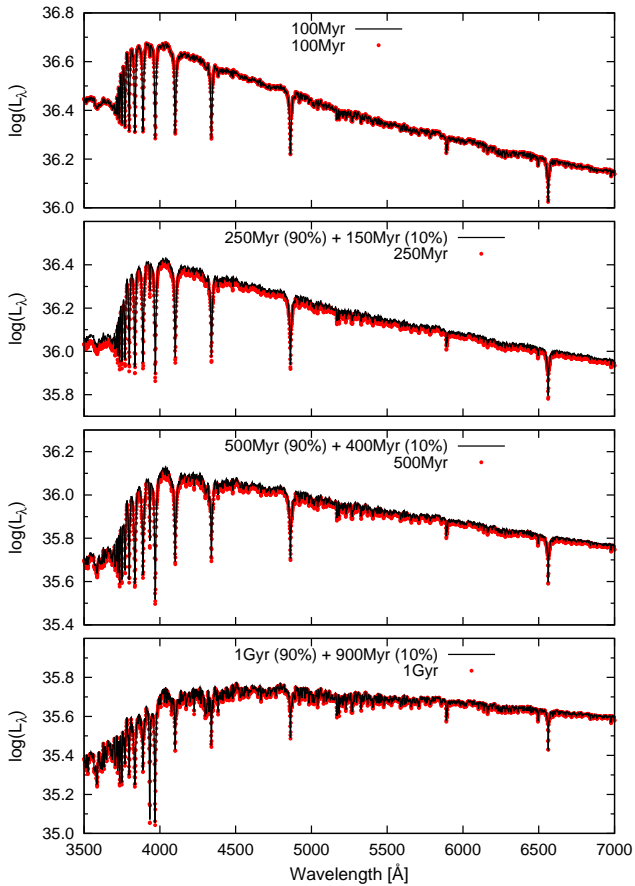} 
 \caption{Spectra of a solar metallicity star cluster produced from the \citet{Bruzual03} models. The red points show the spectrum of a cluster with only one burst of star formation. The black lines show the spectra of a similar cluster but but with an additional burst of star formation 100~Myr after the initial burst. Similarly to the cluster colors plotted in Figure 1, the second generation is taken to be 10$\%$ of the total cluster mass. The cluster is shown at 100~Myr (when only one stellar population exists), 250~Myr, 500~Myr and 1~Gyr (from top to bottom, respectively). } 
 \label{fig:bc03_seds} 
\end{figure}

\begin{figure}
 \centering
 \includegraphics[height=120mm,angle=0]{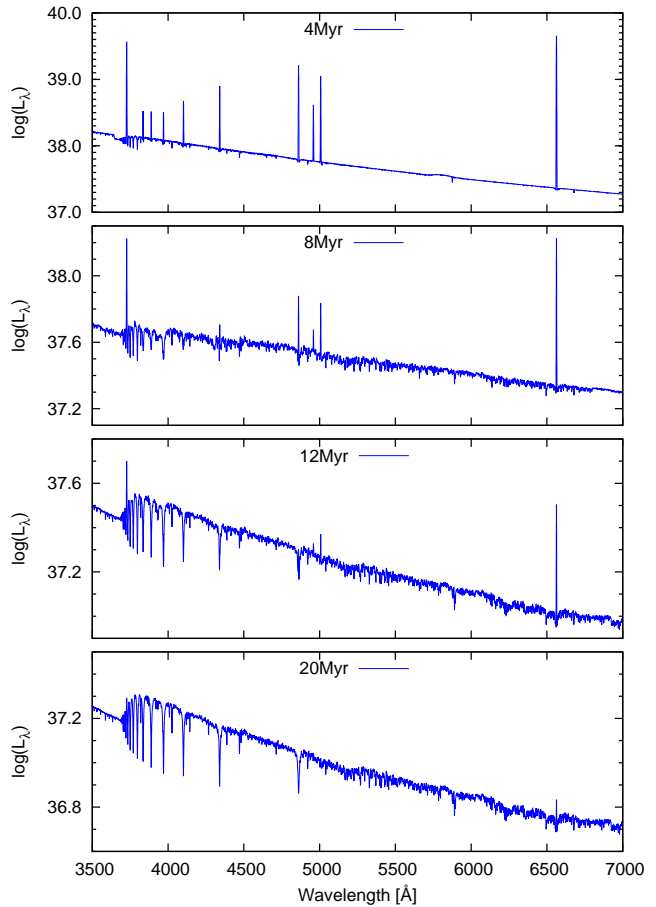} 
 \caption{Spectra of a SSP star cluster with solar metallicity shown at 4, 8, 12 and 20~Myr (from top to bottom). These spectra are produced using the Starburst99 high resolution models with the addition of nebula emission lines (see text for details). The spectrum of the 4~Myr cluster is dominated by strong nebular emission lines. At 8~Myr these lines are weaker, but still clearly detectable. However, by 20~Myr the evolving stellar population produces very few ionizing photons and the spectrum is dominated by the stellar continuum, featuring strong absorption lines. } 
 \label{fig:Sb99_seds} 
\end{figure}

We conclude that the identification of clusters with colors in regions 1 or 2 may be suggestive that they host multiple stellar populations. However, complications due to reddening and model/photometric accuracy mean that this combination of U-B and V-I colors on their own can not conclusively demonstrate the presence of multiple populations in a cluster. A better understanding of the variations between different stellar population models and the addition of photometry extending deeper in to the ultraviolet and infrared wavelengths may help to draw more conclusive evidence based on broadband photometry. While an interesting and important prospect for the future, such work is beyond the scope of the current paper. Finally, we note a caveat to these color considerations is that the second population is proposed to be He enhanced. Such an enhancement may influence the integrated cluster colors, by making the second generation stars bluer. Unfortunately, stellar population models which include He enhancement are still in their infancy and not yet complete enough to consider in a similar fashion. The \citet{Bruzual03} models also exclude the contribution of nebular emission to the cluster's broadband colors. Investigation of the Starburst99 models \citep[which include nebular emission,][]{Leitherer99} suggests that such emission can influence the cluster colors slightly at very young ages but should not significantly effect the general signatures of multiple populations discussed.

\subsection{Optical spectroscopy} 
\label{sec:signatures_spectra}

\begin{figure*}
 \centering
 \includegraphics[height=170mm,angle=270]{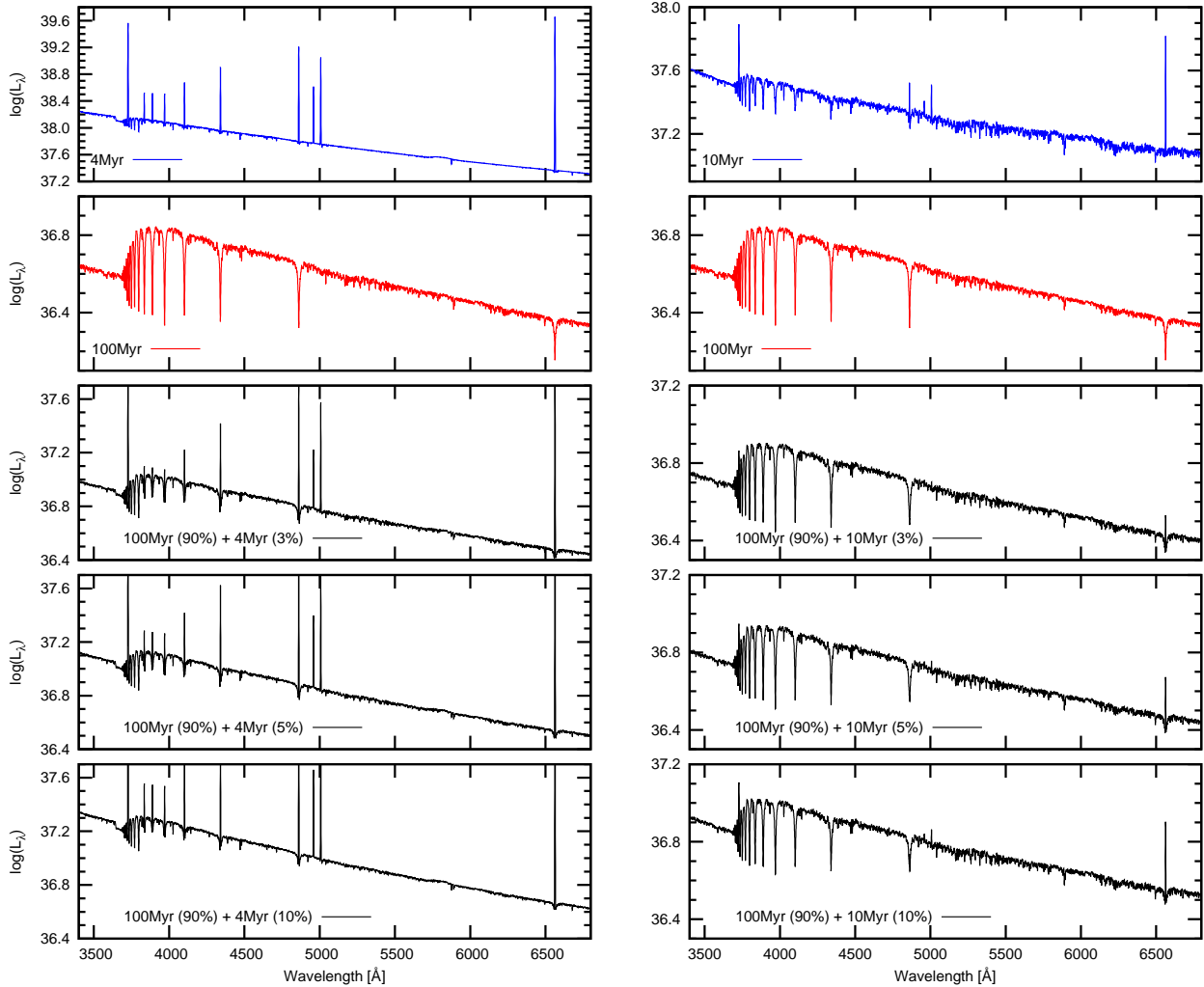} 
 \caption{Starburst99 models with emission lines for solar metallicity SSPs at 4 (blue, left), 10 (blue, right) and 100~Myr (red). The lower panels show the spectra expected from a cluster composed of a 100~Myr stellar population and an additional minority population of 4 (left) and 10~Myr (right). We consider the initial mass ratio of the second population to initial population be 3, 5 and 10$\%$ (middle to bottom, respectively). In all cases, the absorption features of the old population and the emission lines from young population can be observed. } 
 \label{fig:Sb99_98_6} 
\end{figure*}

Integrated spectroscopy of star clusters is available/ obtainable for many nearby galaxies. In Figure \ref{fig:bc03_seds}, we show the predicted spectra of a star cluster produced from the stellar population models of \citet{Bruzual03}. The red circle points show the spectra of a solar metallicity SSP at four different ages -- 100~Myr, 250~Myr, 500~Myr and 1~Gyr. Overlaid are the spectra of the same star cluster at these times, but with an additional burst of star formation at 100~Myr. Again, we assume that this second population of stars comprises 10$\%$ of the total mass of the cluster. In the 250~Myr old cluster, the multiple population cluster is slightly brighter, particularly at short wavelengths. However, given typical errors on observed spectra at these wavelengths and model dependent variations, such a spectrum is adequately fit to a slightly lower mass and younger single stellar population model. Therefore, as observed in the colors presented in Section \ref{sec:signatures_color}, we find that the spectra of a multiple stellar population cluster at these wavelengths and ages can be reasonably well represented as a SSP. More detailed consideration of the relative absorption lines strengths, rather than just considering the entire spectrum with no weighting, may allow multiple populations to be distinguished \citep[such as the methods used in investigating galaxies, e.g.][]{Vergely02,Panter03,Mathis06}. These methods require high signal to noise spectra, reliable stellar population models and a detailed understanding model limitations and fitting procedures. Such considerations are beyond the scope of this paper. Instead, we focus on a clearer signature of multiple populations, discussed below. 

At ages of a few Myr the emission from a cluster features both nebular and stellar emission. This is because the young massive stars in the cluster can photoionize surrounding gas, resulting in a nebula that significantly influences a cluster's emission. Many of the commonly used stellar population models \citep[e.g.][]{Bruzual03,Maraston05} only consider the stellar emission from a cluster. Currently, only two models include both stellar and nebular emission. The GALEV models \citep{Kotulla09} are the only models which currently predict spectra for a stellar population with nebular continuum and emission lines included. However, these models focused on the effects of nebular emission on photometric colors and currently produce spectra with a resolution of only $\sim$20\Ang, lower than most observed spectra. 

The other model to consider the ionization of gas by a stellar population is the Starburst99 model. This code predicts the colors and spectra for a given mass, age and metallicity. While high resolution spectra ($\sim$2.5\Ang) can be produced by this code, they only include the continuum contribution from any ionized gas, excluding any emission lines that would be produced. However, Starburst99 does compute the luminosity of the \Hb emission line. We use this to predict the luminosity of all of the Balmer lines using the relative line strengths quoted by \citet[][assuming Case B, a gas density of 10$^{2}$~cm$^{-3}$ and T=10$^{4}$~K]{Osterbrock89}. Additionally, we estimated the strength of the other prominent lines expected, the \oii and \oiii lines. The combined strength of these oxygen lines was estimated by assuming solar metallicity and using the empirical relationship between metallicity and R$_{23}$~=~(\oiil+\oiiill)/\Hb, as presented in \citet{Kewley02}. The ratio of \oii to \oiii was estimated as \oii=2.5\oiii, assuming a typical ionization parameter \citep{Kewley02}. Finally, we fixed \oiiil~=~3$\times$[O~{\sc iii}]$\lambda$~4959. This last relation is required by atomic physics, but it should be noted that some of the other ratios can vary. For our illustrative purposed, we have simply used typical values for these ratios. The ionized gas in HII regions generally has a low velocity dispersion and would produce unresolved lines in spectra with the resolution of these models. We therefore added these emission lines to the Starburst99 synthetic spectra as Gaussian functions with FWHM=2.5\Ang, similar to the spectral resolution of the models. 

Figure \ref{fig:Sb99_seds} shows the high resolution Starburst99 model spectra for a solar metallicity SSP with emission lines included. We show the model at four ages (4, 8, 12 and 20~Myr). Massive, young stars in the 4~Myr stellar population produce many ionizing photons and the cluster's spectrum features strong emission lines. By 12~Myr the nebular lines are much weaker due to the evolution of the most massive, hottest stars and subsequent reduction in the number of ionizing photons. However, ionizing photons are still present and the stronger lines are still clearly observable above the cluster's continuum. By 20~Myr the nebular emission is extremely weak and the spectrum is dominated by stellar emission. 

In Figure \ref{fig:Sb99_98_6} we use these models to show the spectrum produced by adding a minority second stellar population of 4~Myr (left, blue) and 10~Myr (right, blue) to a 100~Myr old population (red). The initial ratio of first to second generation stars at these young ages is still poorly understood. We therefore consider three different cases where the minority (younger) population comprises 3, 5 and 10$\%$ of the cluster's mass. In all cases the stellar like spectrum from the older stellar population, featuring broad absorption lines, can be observed in the spectra of the multiple population clusters. In addition to this, the narrow emission lines produced by the younger second population can also be observed. For the 4~Myr case, the emission lines are extremely strong and all of them can be observed in the multiple population cluster, even when the younger population comprises only 3$\%$ of the total population. When the second population reaches 10~Myr, the nebular emission is much weaker. However, the strongest emission line (\Ha) is still observable above the stellar like continuum. Therefore, assuming that the second population in a two population cluster is younger than $\sim$10~Myr and comprises more than a few $\%$ of the total cluster population, the spectrum of the multiple population cluster should feature both strong absorption and emission lines. 

\begin{figure}
 \centering
 \includegraphics[height=120mm,angle=0]{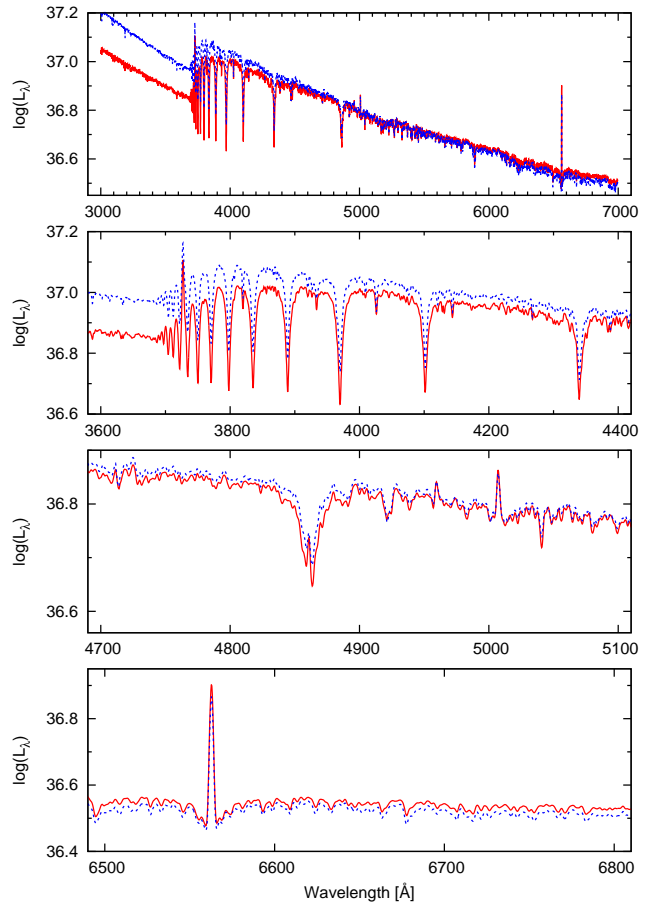} 
 \caption{Starburst99 spectra of a SSP cluster at 14~Myr (blue dotted line) and a multiple stellar population cluster with 90$\%$ of its stars at 100~Myr and 10$\%$ at 10~Myr (red solid line). The full cluster spectrum is shown in the top panel and lower panels show zoomed in sections. The two models have similar absorption and emission lines. However the continuum of the multiple population is dominated by the more numerous older population of stars and is much redder than the 14~Myr cluster. } 
 \label{fig:comp_ssp_msp} 
\end{figure}

Could a SSP produce such a spectrum? The ratio of emission to absorption line strengths in the spectrum of a young cluster ($<$10~Myr) would be too high to fit such a spectrum. Similarly, the emission lines in the spectrum of the multiple population cluster could not be produced by a stellar population older than $\sim$20~Myr. We note that older populations can potentially feature emission lines from planetary nebulae. However, these are known to be extremely rare in globular clusters \citep{Peacock12a} and would struggle to produce both the predicted emission line flux and the correct ratio of \Ha/\oiii. There is a small range of ages around 12-16~Myr where a SSP cluster may feature absorption and emission lines. Such a cluster could not fit the strong emission lines in the spectrum of a multiple stellar population cluster with a second generation younger than $\sim$8~Myr. However, it could have a similar ratio of absorption to emission lines as a multiple stellar population with an older second generation ($\sim$10~Myr). In Figure \ref{fig:comp_ssp_msp}, we compare two such models (scaled to the same flux at 5000\Ang) -- a single stellar population at 14~Myr and a multiple stellar population with 90$\%$ of the cluster mass in stars that are 100~Myr old and the other 10$\%$ in stars that are 10~Myr old. It can be seen that the two models can produce similar strength emission and absorption lines. However, the continuum of the multiple stellar population (dominated by the 100~Myr population) could not be well represented by the younger 14~Myr population -- with the 14~Myr population producing too much flux at short wavelengths and too little flux at long wavelengths. This is analogous to the gap in region 1 of the color-color plot shown in Figure \ref{fig:bc03_colors}. Such a discrepancy can be used to distinguish between these intermediate age, single stellar population clusters, and a multiple stellar population cluster. We conclude that, unlike at older ages, the spectrum of a multiple stellar population cluster, where the younger population is $<$10~Myr, can not be well represented by a single stellar population cluster with either a young, old or intermediate age. Consideration of these models therefore predicts a key period when a cluster comprised of multiple stellar populations can be clearly identified based on its integrated spectroscopy: 

{\it When the second stellar population in a star cluster is younger than $\sim$10~Myr, the presence of multiple stellar populations should be identifiable in integrated spectroscopy due to a stellar continuum that indicates an older stellar population in addition to emission lines, that require an additional, very young population.} 

When the younger population is older than $\sim$10~Myr, any nebular emission is very faint and the key spectral features of the two populations are similar. This makes clear identification of more than one stellar population in the observed integrated emission much more challenging and more model dependent. 

\subsection{Probability of observing such signatures} 
\label{sec:stats}

Given the above considerations, we believe that the clearest detectable signature of multiple stellar populations in an unresolved cluster occurs when the youngest population is $<$10~Myr. The presence of emission lines from this population in addition to the older cluster spectrum should be detectable in spectroscopy of the cluster. This requires an age spread of $>$20~Myr between the populations (so that the first population is dominated by absorption features). This would be expected if the second generation gas is enriched via winds from AGB stars, as has been proposed.

We can estimate the likelihood of detecting a cluster during the period when the multiple stellar populations can be detected using this method. To do this, we assume that all clusters have two stellar populations (additional bursts of star formation will tend to increase the likelihood of detection by increasing the number of epochs when the young population produces nebular emission). The time between the two bursts of star formation (and whether this is universal to all clusters) is currently unknown. We therefore assume that the second population has an equal probability of being formed at any epoch in the range permissible by the leading multiple population formationary theory, that invoking enrichment from AGB stars -- 40-200~Myr. Finally, using this method, the second generation is detected via its associated nebular emission and is therefore detectable for a period of $\sim$10~Myr. Given these considerations, we predict that there is a 6.5$\%$ chance of observing a random star cluster (with an age of 40-200~Myr) during the period when its second stellar population is younger than 10~Myr and thus detectable. Therefore, one would expect 1 in every 15 such clusters to exhibit such a signature and there would a $>$99$\%$ chance of detecting such a signature from at least one cluster in a sample of 70 clusters. Spectroscopic samples of this size are potentially observable in nearby galaxies. Candidates for clusters that are currently at this interesting period in their evolution could be identified in photometric surveys as having colors similar to region 1 in Figure \ref{fig:bc03_colors}.

\section{Observing multiple populations in massive young clusters: a pilot study} 
\label{sec:obs}

\subsection{Young massive clusters in the Antennae galaxies}
\label{sec:antennae}

The interacting Antennae galaxies provide a rich source of young star clusters. These clusters have been the focus of many photometric and spectroscopic studies. In this section, we discuss a search for such signatures of multiple stellar populations in published spectroscopy of these clusters. 

A spectroscopic survey of 15 of the massive young star clusters in the Antennae galaxies was presented by \citet{Bastian09}. This survey confirmed that these clusters have ages in the range of a few Myr to a few hundred Myr and hence lie in the range where signatures of of multiple populations can be detected. We briefly review the data used by this study, but for a more detailed discussion we refer the reader to \citet{Bastian09}. The spectra of these clusters were obtained using the multi-object mode of the Gemini Multi Object Spectrograph (GMOS) on Gemini-North under the program GN-2003A-Q-33 (PI: Gelys Trancho). Clusters were targeted based on preimaging supplemented by \hst observations to confirm their cluster status. Masks with slit widths of 0.75$\arcsec$ were used to observe the clusters using the B600 grating, with a resolution R=1688 at 4610\Ang. These spectra were presented and discussed in \citet{Bastian09} and we do not repeat their analysis. However, we note that seven of the clusters targeted show emission line spectra consistent with young SSPs of a few Myr and seven clusters have spectra that are dominated by absorption lines and consistent with SSPs of a few 100~Myr \citep[see figure 3 of][]{Bastian09}. Interestingly, the other cluster targeted (T111) shows both absorption and emission lines. While the absorption features can be well fit by a SSP of 80~Myr, the strong emission lines can not be produced by such a population. This source therefore exhibits the signature expected from multiple stellar populations. 

\subsection{The Antennae star cluster T111} 

\begin{figure}
 \centering
 \includegraphics[height=86mm,angle=270]{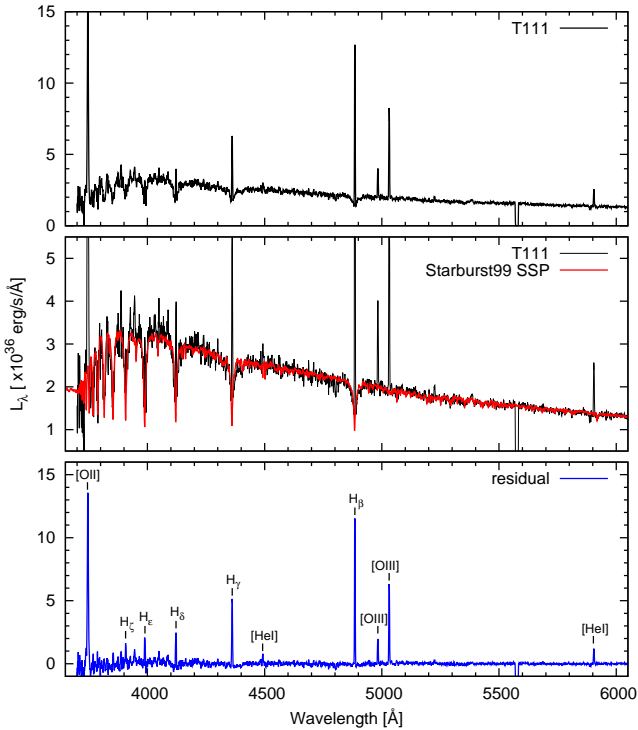} 
 \caption{GMOS spectrum of the cluster T111 (black line, top). The middle panel compares this to an 80~Myr, solar metallicity SSP model from the Starburst99 code (red line). The bottom panel shows the residual after subtraction of the 80~Myr model. The model provides a good representation of the absorption line spectrum. However, strong emission lines that are inconsistent with such a model are also present. Marked in this lower panel are the (redshifted) wavelengths of the Balmer series, ionized oxygen \oii and \oiii, and neutral He [He~{\sc i}] lines. }
 \label{fig:T111_spectrum} 
\end{figure}

To further investigate the intriguing cluster T111, we obtained the raw spectra from the Gemini Science Archive\footnote{http://www3.cadc-ccda.hia-iha.nrc-cnrc.gc.ca/gsa/}. The spectra were reduced using the standard {\sc iraf gemini} packages. The raw spectra were first prepared for the {\sc gemini} packages using {\sc gprepare} and normalized flatfield images were produced using {\sc gsflat}. The edges of each slit on the image were found from the combined flatfield images using {\sc gscut}. The task {\sc gsreduce} was then used to subtract off the bias image (provided by the Gemini archive), clean for cosmic ray hits, mosaic the three detectors together, apply the flatfield correction and cut the image (containing all of the multi-object spectra) in to individual spectra. A wavelength calibration was applied to the images based on associated CuAr arc images, using the task {\sc gswavelength}. This produced the final set of six 2D spectral images of T111. The spectra were then located, traced and extracted from these images using the {\sc iraf/noao apall} task with an extraction width of 9 pixels ($\sim$1.3\arcsec). Before combining, the resulting spectra were then corrected for slit losses due to atmospheric differential refraction. This was required because the spectra were generally not taken at the parallactic angle -- although the choice of the (fixed) slit angle minimized this effect. This correction was based on the method of \citet{Filippenko82} for our $0.75\arcsec \times 7.0\arcsec$ slit, measured airmasses of 1.3-1.5, seeing of 0.8-1.0\arcsec and angles to the parellactic angle of 3.5, 5.0, 6.6, 18.5, 24.0, 43.1$^{\circ}$. The maximum correction to the flux at 3700\Ang, relative to 5000\Ang, was a factor 1.5. 

The spectrum of T111 is known to be well represented by a SSP model of solar metallicity and an age of 80~Myr \citep{Bastian09}. We use such a model to apply an approximate flux calibration to the spectrum. This was done by fitting a smooth function to the ratio of the model to the GMOS spectrum and scaling the spectrum using this function. The final spectrum of T111 is shown in the top panel of Figure \ref{fig:T111_spectrum}. The spectrum is in excellent agreement with that presented by \citet{Bastian09}. The middle panel of this figure compares this spectrum with a Starburst99 SSP model with solar metallicity and an age of 80~Myr. The bottom panel of Figure \ref{fig:T111_spectrum} (blue) shows the residual emission after subtracting this model from the spectrum. While the model provides a good representation of the cluster's spectrum, strong emission lines are also present. These are nebular emission lines with the Balmer series, ionized oxygen \oiil and \oiiill, and neutral He [He~{\sc i}]~$\lambda\lambda$4471,5876 lines all clearly visible. The observed wavelengths of these lines are in excellent agreement with the expected rest frame wavelengths, assuming a velocity of 1383~kms$^{-1}$ (as marked in Figure \ref{fig:T111_spectrum}). The luminosity of these emission lines was measured using {\sc splot}. They are well represented by a Gaussian with FWHM=3.5\Ang, consistent with the spectral resolution of the data, as expected for a typical HII region. 

\subsubsection{The two components of T111: a binary cluster?}

\begin{figure}
 \centering
 \includegraphics[width=86mm,angle=0]{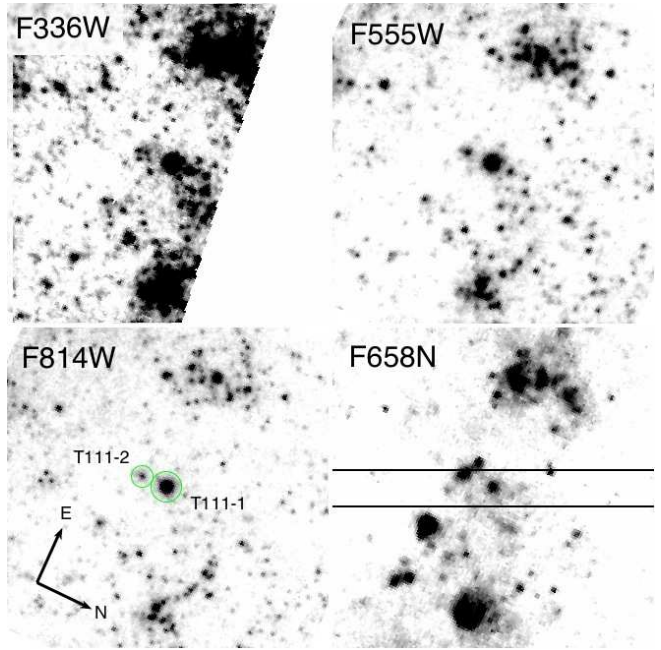} 
 \caption{\hst WFC3/UVIS images of T111 in the broadband F336W (U), F555W (V) and F814W (I) filters. The bottom right panel shows the ACS/WFC narrowband F658N (\Ha) image of this region and the location of the GMOS slit. The bottom left figure indicates the orientation of all of these images and highlights the two clusters, T111-1 and T111-2. }
 \label{fig:T111_images} 
\end{figure}

Multiwavelength archival \hst photometry of this cluster is available from several surveys of the Antennae. We utilized the most recent optical survey of the Antennae (obtained under PID 11577, P.I. Whitmore). This provides broadband F336W, F555W, F625W and F814W photometry (from WFC3/UVIS observations), broadband F435W and narrowband F658N photometry (from ACS/WFC observations). We obtained the pipeline reduced and combined images of the cluster in these different filters from the Hubble Legacy Archive. In these images, the cluster was found to be offset by 2$\arcsec$ from the location quoted by \citet{Bastian09}, lying in our images at $\alpha$=12:01:53.3, $\delta$=-18:51:37.8. This is likely due to an error in the absolute astrometry of either our or their images. However, the location we take for the cluster was confirmed to be the object targeted by the Gemini spectroscopy based on the image presented in figure 4 of \citet{Bastian09}. 

The narrowband F658N \hst photometry (covering \Ha at the velocity of this cluster) showed that T111 does not exhibit strong excess emission in the \Ha filter, as would be expected given the strong Balmer emission lines observed in its spectrum. However, a source very close to the targeted cluster was found to have a strong \Ha excess. At only 0.62$\arcsec$ from the cluster, this additional source would be blended with the cluster emission in the Gemini observations and included in the 0.75$\arcsec$ slit (see Figure \ref{fig:T111_images}). We believe it to be extremely likely that this second source provides the optical emission lines observed in T111's spectrum. Both of the sources are resolved in these images and are believed to be two distinct clusters with absolute V-band magnitudes of -10.6$\pm$0.1 and -8.6$\pm$0.2 \citep[assuming a distance of 22~Mpc,][]{Schweizer08}. The photometry of these two clusters was performed using {\sc iraf/apphot}. For the brighter, older cluster (T111-1) we used an aperture with a radius of 0.3\arcsec. For the fainter, younger cluster (T111-2), we used a smaller 0.2\arcsec aperture because of contamination from nearby sources. Comparing the colors obtained for these two clusters with the SSP models of \citet{Bruzual03}, we find that T111-1 is well represented by an 80~Myr old, solar metallicity cluster with mass 40$\pm$4$\times$10$^{4}{\rm M_{\odot}}$ and T111-2 by an 8~Myr old, solar metallicity cluster with mass 10$\pm$2$\times$10$^{3}{\rm M_{\odot}}$. 

The \hst photometry of the T111 system therefore suggests that the old and young components observed in the GMOS spectrum are likely due to two distinct clusters. We conclude that T111 is not an example of a single young cluster hosting multiple populations. However, the proximity of these two clusters to each other (0.62\arcsec, corresponding to a projected separation of 59~pc at the distance of the Antennae galaxies) raises the question of whether their coalescence could result in a single cluster with multiple populations. Binary clusters have been proposed in other galaxies, such as the LMC \citep{Dieball02, Mucciarelli12} and NGC~5128 \citep{Minniti04} and models show that binary clusters can be formed during high velocity cloud interactions, as would be expected in the Antennae \citep{Fujimoto97, Bekki04}. The coalescence of star clusters potentially provides a method of producing a single globular cluster hosting multiple stellar populations -- although such a scenario struggles to produce the observed abundance patterns (such as the similar [Fe/H] and the Na-O anticorrelation). 

The close alignment of the Balmer emission and absorption lines in Figure \ref{fig:T111_spectrum} demonstrates that T111-1 and T111-2 must have similar radial velocities. Using the brightest emission lines in the spectrum (\Hb, H$_{\gamma}$, H$_{\delta}$, \oiiill and \oiil) we measure T111-2's radial velocity, $v=1383\pm 4$~kms$^{-1}$ (where the error is estimated as the standard deviation of the velocities measured for each line). The radial velocity of T111-1 was measured from the absorption lines using the cross-correlation {\sc iraf} routine {\sc fxcor}. For our reference spectrum we used the high resolution Starburst99 SSP model for a solar metallicity, 80~Myr old star cluster. The cross correlation was run over the regions of the spectrum that were unaffected by the emission lines. We experimented with several different wavelength ranges, achieving consistent results. The most accurate correlation was produced by using most of the spectrum, from 3900-6100\Ang (excluding the emission lines), for which the cross correlation resulted in a radial velocity and associated error of $1391\pm 17$~kms$^{-1}$. Because the stellar population models vary slightly in their resulting spectra, we also compared our spectrum to that of a similar cluster produced using the updated models of \citet{Maraston05, Maraston11}. Excellent agreement was found between the two models with the cross correlations producing velocity differences of only a few kms$^{-1}$. 

For T111-1 and T111-2 to be gravitationally bound, we require that the potential energy is greater than the kinetic energy, or:

\begin{equation}
 \label{equ:e1}
 \frac{M_{sys}v^{2}} {2} < \frac{GM_{sys}^{2}} {R}
\end{equation}

Here $M_{sys}$ is the total mass of the two clusters, $R$ is their separation and $v$ is their relative velocity. Unfortunately, it is only possible to measure the line of sight velocity ($v_{los}$) and the projected separation ($R_{p}$) between the two clusters. We substitute these parameters in to Equation \ref{equ:e1} by following the method of \citet{Davis95} and defining the system to be at an angle $\alpha$ to the sky: 

\begin{equation}
 \label{equ:e2}
 \frac{v_{los}^{2}R_{p}} {2GM_{sys}} < sin^{2} \alpha cos \alpha \quad (0<\alpha<90^{\circ})
\end{equation}

For the T111 system, we have measured M$_{1}$=10$^{5.6}{\rm M_{\odot}}$, M$_{2}$=10$^{4.0}{\rm M_{\odot}}$ and $R_{p}$=59~pc, but the angle $\alpha$ is unknown. However, for most angles ($10<\alpha<80^{\circ}$), we find that the maximum permissible $v_{los}$ between two bound clusters is $3.1<v_{los,esc}<4.8$~kms$^{-1}$ (depending on $\alpha$). The measured line of sight velocity difference between T111-1 and T111-2 is consistent with the clusters being bound. However, the accuracy of the radial velocity estimates is too low to be conclusive. 

\subsubsection{Abundance Analysis of T111-1 and T111-2}

The spectrum of T111-1 suggests that the metallicity of its stellar population is around solar, 0.9$\pm$0.3Z$_{\odot}$ \citep{Bastian09}. The metallicity of the younger cluster, T111-2, can also be estimated based on its emission line strengths. To do this we use the R$_{23}$ diagnostic, defined as:

\begin{equation}
 \label{equ:e3}
 \rm
 R_{23}=\frac{[O~\textsc{ii}]\lambda\lambda~3726, 3729+[O~\textsc{iii}]\lambda\lambda~4959, 5007} {H_{\beta}}
\end{equation}

From the continuum subtracted spectrum (presented in the bottom panel of Figure \ref{fig:T111_spectrum}), we estimate these line fluxes as: \Hb=4.6$\times$10$^{37}$ergs$^{-1}$; \oiiill=3.2$\times$10$^{37}$ergs$^{-1}$; \oiill=5.0$\times$10$^{37}$ergs$^{-1}$. This implies log(R$_{23}$)=0.25 for T111-2. Based on the noise of the continuum, we estimate that the formal errors on these strong emission lines are $<$1$\%$. The uncertainties are likely to be larger than this due to the flux calibration of the spectrum, particularly for the ratio of \oii to the redder lines. However, the uncertainty on the line fluxes are believed to be smaller than the large systematic error associated with using these lines to find [Fe/H] and the He abundance, discussed below. 

Empirical relationships between R$_{23}$ and log(O/H) have been extensively studied, although significant scatter is known to exist producing systematic errors of 0.2-0.5~dex \citep[][and references therein]{Kobulnicky04}. We use the relationship between R$_{23}$ and log(O/H) that is presented in equation 16 of \citet{Kobulnicky04}. This is the average of the high metallicity relationships presented by \citet{McGaugh91} and \citet{Kewley02}. Using this relation we find that log(O/H)+12=9.0 which, assuming solar abundances, implies a metallicity of 1.2Z$_{\odot}$. This is higher than that of T111-1, but consistent given the predicted uncertainties. 

In addition to the \oii, \oiii, and Balmer emission lines, [He~{\sc i}]$\lambda$~4471 and [He~{\sc i}]$\lambda$~5876 emission lines are clearly seen in the spectrum (Figure \ref{fig:T111_spectrum}). In principle, these lines can be used to constrain the He abundance in the T111-2 nebula. \citet{Benjamin99} present the predicted emission coefficient, $4\pi f_{line}/n_{e}n_{{\rm He}}$, as a function of the gas density and temperature. Based on these predictions (for $n_{e}$=10$^{2}$~cm$^{-3}$ and T=10$^{4}$~K) and using the observed emission line flux ratio, $f_{{\rm He,4472}}/f_{{\rm H_{\beta}}}$=0.075, we can estimate the He number abundance as: 

\begin{equation}
 \label{equ:e3}
 \frac{n_{{\rm He}}} { n_{{\rm H_{\beta}}}} = \frac{1.23\times10^{-25}}{6.16\times10^{-26}} \frac{f_{{\rm He,4472}}}{f_{{\rm H_{\beta}}}} = 0.15
\end{equation}
\\
This corresponds to a He mass fraction, Y=0.35 (assuming a metal fraction of 0.02). Such high He fractions are proposed in the second generation stars of globular clusters. However, we caution that the density and temperature of the gas that is assumed in this calculation are not very well constrained. While the result is relatively insensitive to higher temperatures at this density, at higher densities and temperatures the predicted line ratios are consistent with primordial abundances. 

\subsubsection{The nature and future of T111}

The stellar populations in the two T111 clusters are consistent with the multiple populations seen in the Galactic globular clusters: they have similar metallicities; an age difference of 70~Myr; and there is even the suggestion that the younger cluster is He enhanced. As demonstrated above, their proximity and similar velocities are consistent with the clusters being bound. The coalescence of clusters like T111-1 and T111-2 may play an important role in the history of at least some star clusters. The resulting mixed population could produce a globular like cluster with multiple populations, similar to those observed in old globular clusters. Mergers of cluster systems have been proposed in environments like the Antennae \citep[e.g.][]{Kroupa98}. Additionally, mergers have been proposed as a possible explanation for observations of some star clusters -- such as the extended and massive Milky Way halo globular cluster NGC~2419 \citep{Bruns11}, the `faint fuzzy' star clusters observed in the halos of other galaxies \citep{Bruns09, Assmann11} and the observed mass segregation in some Galactic open clusters \citep[e.g.][]{McMillan07, Moeckel09}. 

The coalescence of sub-clusters clusters may be important in the history of at least some clusters. However, the T111 system is clearly different than the prototypical multiple population cluster that is proposed by most current models. These models generally consider isolated clusters where the second populations form in the center of the clusters (though in reality the situation is likely more complex). Identification of such multiple populations, within an individual cluster, may be possible in spectroscopic surveys of more clusters in the age range of 10s-100s of Myr.

\section{Conclusions}

We have considered the integrated emission from star clusters composed of single and multiple stellar populations. The stellar population models of \citet{Bruzual03} identify two periods in a cluster's lifetime when the optical U-B and V-I colors of a multiple stellar population are significantly different from the colors of a SSP. However, we note that observational uncertainties, reddening and uncertainties in the metallicity of a cluster make it difficult to confirm the presence of multiple populations in a cluster based on these two optical colors alone. Additionally, we note that there are significant variations in the colors predicted by different stellar population models at young ages. This further complicates the analysis of cluster colors. We note that better understanding of variations between models and the extension of this work to the ultraviolet and near-infrared may help to identify multiple population clusters based on broadband colors in the future.

We also find that it is challenging to identify multiple stellar populations (with small age differences of order 100~Myr) in the spectra of old clusters. This is because the spectral features of the different populations tend to be similar and their optical spectra are adequately represented by SSP models. While future work may potentially be able to identify multiple populations based on inconsistent line ratios for a SSP, we note that such studies will require well understood stellar population models and high signal to noise spectra. 

We have identified a strong and clear signal of multiple stellar populations in a young cluster's spectrum. This occurs when the youngest stellar population is $\lesssim$10~Myr old. The young massive stars present at these young ages will produce strong nebular emission. Such emission can not be produced by star clusters with ages $\lesssim$16~Myr. Therefore, a cluster featuring both strong absorption lines (that require an older stellar population) and narrow nebular emission lines (that require a young population of stars) could only be produced by two distinct stellar populations. 

As a preliminary search, we have looked for such a cluster in published spectroscopy of 15 massive young star clusters in the Antennae. Seven of these clusters show emission lines and are consistent with young stellar populations and seven show absorption lines that are consistent with older ages. Interestingly, the other cluster shows both strong absorption and emission lines (suggestive of two populations with ages of 8~Myr and 80~Myr). We show that this emission comes from two neighboring sub-clusters that are consistent with being bound. The properties of these two clusters suggest that their coalescence could produce a single multiple population cluster that is similar to that of the Galactic globular clusters. 

If all massive star clusters contain multiple stellar populations, as appears to be the case, then spectroscopy of more clusters in the 40-200~Myr range should identify a cluster with its second population younger than 10~Myr and hence showing these dual spectral features consistent with multiple stellar populations. Identification of these clusters is clearly extremely desirable as it can help to constrain the still poorly understood formationary theory of star clusters in the light of multiple stellar populations.

\section*{Acknowledgments}
We thank the anonymous referee for their detailed and constructive comments which were beneficial to the final paper. We also thank Enrico Vesperini, Arunav Kundu and Tom Maccarone for helpful discussions. MBP and SEZ acknowledge partial support from the NASA grant NNX08AJ60G.

Based on observations obtained at the Gemini Observatory (acquired through the Gemini Science Archive), which is operated by the Association of Universities for Research in Astronomy, Inc., under a cooperative agreement with the NSF on behalf of the Gemini partnership: the National Science Foundation (United States), the National Research Council (Canada), CONICYT (Chile), the Australian Research Council (Australia), Ministério da Ciência, Tecnologia e Inovação (Brazil) and Ministerio de Ciencia, Tecnología e Innovación Productiva (Argentina). 

Based on observations made with the NASA/ESA Hubble Space Telescope, and obtained from the Hubble Legacy Archive, which is a collaboration between the Space Telescope Science Institute (STScI/NASA), the Space Telescope European Coordinating Facility (ST-ECF/ESA) and the Canadian Astronomy Data Centre (CADC/NRC/CSA).

\label{lastpage}


\begin{thebibliography}{55}
\expandafter\ifx\csname natexlab\endcsname\relax\def\natexlab#1{#1}\fi

\bibitem[{{Ashman} \& {Zepf}(1998)}]{Ashman98}
{Ashman}, K.~M. \& {Zepf}, S.~E. 1998, {Globular Cluster Systems}

\bibitem[{{Assmann} {et~al.}(2011){Assmann}, {Wilkinson}, {Fellhauer}, \&
  {Smith}}]{Assmann11}
{Assmann}, P., {Wilkinson}, M.~I., {Fellhauer}, M., \& {Smith}, R. 2011,
  \mnras, 413, 2606

\bibitem[{{Bastian} {et~al.}(2009){Bastian}, {Trancho}, {Konstantopoulos}, \&
  {Miller}}]{Bastian09}
{Bastian}, N., {Trancho}, G., {Konstantopoulos}, I.~S., \& {Miller}, B.~W.
  2009, \apj, 701, 607

\bibitem[{{Bekki}(2011)}]{Bekki11}
{Bekki}, K. 2011, \mnras, 412, 2241

\bibitem[{{Bekki} {et~al.}(2004){Bekki}, {Beasley}, {Forbes}, \&
  {Couch}}]{Bekki04}
{Bekki}, K., {Beasley}, M.~A., {Forbes}, D.~A., \& {Couch}, W.~J. 2004, \apj,
  602, 730

\bibitem[{{Bekki} \& {Norris}(2006)}]{Bekki06}
{Bekki}, K. \& {Norris}, J.~E. 2006, \apjl, 637, L109

\bibitem[{{Benjamin} {et~al.}(1999){Benjamin}, {Skillman}, \&
  {Smits}}]{Benjamin99}
{Benjamin}, R.~A., {Skillman}, E.~D., \& {Smits}, D.~P. 1999, \apj, 514, 307

\bibitem[{{Bragaglia} {et~al.}(2010){Bragaglia}, {Carretta}, {Gratton},
  {Lucatello}, {Milone}, {Piotto}, {D'Orazi}, {Cassisi}, {Sneden}, \&
  {Bedin}}]{Bragaglia10}
{Bragaglia}, A. {et~al.} 2010, \apjl, 720, L41

\bibitem[{{Br{\"u}ns} \& {Kroupa}(2011)}]{Bruns11}
{Br{\"u}ns}, R.~C. \& {Kroupa}, P. 2011, \apj, 729, 69

\bibitem[{{Br{\"u}ns} {et~al.}(2009){Br{\"u}ns}, {Kroupa}, \&
  {Fellhauer}}]{Bruns09}
{Br{\"u}ns}, R.~C., {Kroupa}, P., \& {Fellhauer}, M. 2009, \apj, 702, 1268

\bibitem[{{Bruzual} \& {Charlot}(2003)}]{Bruzual03}
{Bruzual}, G. \& {Charlot}, S. 2003, MNRAS, 344, 1000

\bibitem[{{Carretta} {et~al.}(2009{\natexlab{a}}){Carretta}, {Bragaglia},
  {Gratton}, {D'Orazi}, \& {Lucatello}}]{Carretta09b}
{Carretta}, E., {Bragaglia}, A., {Gratton}, R., {D'Orazi}, V., \& {Lucatello},
  S. 2009{\natexlab{a}}, \aap, 508, 695

\bibitem[{{Carretta} {et~al.}(2009{\natexlab{b}}){Carretta}, {Bragaglia},
  {Gratton}, {Lucatello}, {Catanzaro}, {Leone}, {Bellazzini}, {Claudi},
  {D'Orazi}, {Momany}, {Ortolani}, {Pancino}, {Piotto}, {Recio-Blanco}, \&
  {Sabbi}}]{Carretta09a}
{Carretta}, E. {et~al.} 2009{\natexlab{b}}, \aap, 505, 117

\bibitem[{{Carretta} {et~al.}(2010){Carretta}, {Bragaglia}, {Gratton},
  {Recio-Blanco}, {Lucatello}, {D'Orazi}, \& {Cassisi}}]{Carretta10}
{Carretta}, E., {Bragaglia}, A., {Gratton}, R.~G., {Recio-Blanco}, A.,
  {Lucatello}, S., {D'Orazi}, V., \& {Cassisi}, S. 2010, \aap, 516, A55

\bibitem[{{Chandar} {et~al.}(2010){Chandar}, {Whitmore}, {Kim}, {Kaleida},
  {Mutchler}, {Calzetti}, {Saha}, {O'Connell}, {Balick}, {Bond}, {Carollo},
  {Disney}, {Dopita}, {Frogel}, {Hall}, {Holtzman}, {Kimble}, {McCarthy},
  {Paresce}, {Silk}, {Trauger}, {Walker}, {Windhorst}, \& {Young}}]{Chandar10}
{Chandar}, R. {et~al.} 2010, \apj, 719, 966

\bibitem[{{Conroy}(2012)}]{Conroy12}
{Conroy}, C. 2012, \apj, 758, 21

\bibitem[{{Conroy} \& {Gunn}(2010)}]{Conroy10}
{Conroy}, C. \& {Gunn}, J.~E. 2010, ApJ, 712, 833

\bibitem[{{Conroy} \& {Spergel}(2011)}]{Conroy11}
{Conroy}, C. \& {Spergel}, D.~N. 2011, \apj, 726, 36

\bibitem[{{D'Antona} \& {Ventura}(2007)}]{DAntona07}
{D'Antona}, F. \& {Ventura}, P. 2007, \mnras, 379, 1431

\bibitem[{{Davis} {et~al.}(1995){Davis}, {Bird}, {Mushotzky}, \&
  {Odewahn}}]{Davis95}
{Davis}, D.~S., {Bird}, C.~M., {Mushotzky}, R.~F., \& {Odewahn}, S.~C. 1995,
  \apj, 440, 48

\bibitem[{{D'Ercole} {et~al.}(2008){D'Ercole}, {Vesperini}, {D'Antona},
  {McMillan}, \& {Recchi}}]{DErcole08}
{D'Ercole}, A., {Vesperini}, E., {D'Antona}, F., {McMillan}, S.~L.~W., \&
  {Recchi}, S. 2008, \mnras, 391, 825

\bibitem[{{Dieball} {et~al.}(2002){Dieball}, {M{\"u}ller}, \&
  {Grebel}}]{Dieball02}
{Dieball}, A., {M{\"u}ller}, H., \& {Grebel}, E.~K. 2002, \aap, 391, 547

\bibitem[{{Filippenko}(1982)}]{Filippenko82}
{Filippenko}, A.~V. 1982, \pasp, 94, 715

\bibitem[{{Fujimoto} \& {Kumai}(1997)}]{Fujimoto97}
{Fujimoto}, M. \& {Kumai}, Y. 1997, \aj, 113, 249

\bibitem[{{Goudfrooij} {et~al.}(2011){Goudfrooij}, {Puzia},
  {Kozhurina-Platais}, \& {Chandar}}]{Goudfrooij11}
{Goudfrooij}, P., {Puzia}, T.~H., {Kozhurina-Platais}, V., \& {Chandar}, R.
  2011, \apj, 737, 3

\bibitem[{{Gratton} {et~al.}(2004){Gratton}, {Sneden}, \&
  {Carretta}}]{Gratton04}
{Gratton}, R., {Sneden}, C., \& {Carretta}, E. 2004, \araa, 42, 385

\bibitem[{{Gratton} {et~al.}(2012){Gratton}, {Carretta}, \&
  {Bragaglia}}]{Gratton12}
{Gratton}, R.~G., {Carretta}, E., \& {Bragaglia}, A. 2012, \aapr, 20, 50

\bibitem[{{Gratton} {et~al.}(2013){Gratton}, {Lucatello}, {Sollima},
  {Carretta}, {Bragaglia}, {Momany}, {D'Orazi}, {Cassisi}, {Pietrinferni}, \&
  {Salaris}}]{Gratton13}
{Gratton}, R.~G. {et~al.} 2013, \aap, 549, A41

\bibitem[{{Kewley} \& {Dopita}(2002)}]{Kewley02}
{Kewley}, L.~J. \& {Dopita}, M.~A. 2002, \apjs, 142, 35

\bibitem[{{Kobulnicky} \& {Kewley}(2004)}]{Kobulnicky04}
{Kobulnicky}, H.~A. \& {Kewley}, L.~J. 2004, \apj, 617, 240

\bibitem[{{Kotulla} {et~al.}(2009){Kotulla}, {Fritze}, {Weilbacher}, \&
  {Anders}}]{Kotulla09}
{Kotulla}, R., {Fritze}, U., {Weilbacher}, P., \& {Anders}, P. 2009, \mnras,
  396, 462

\bibitem[{{Kraft}(1994)}]{Kraft94}
{Kraft}, R.~P. 1994, \pasp, 106, 553

\bibitem[{{Kroupa}(1998)}]{Kroupa98}
{Kroupa}, P. 1998, \mnras, 300, 200

\bibitem[{{Langer} {et~al.}(1993){Langer}, {Hoffman}, \& {Sneden}}]{Langer93}
{Langer}, G.~E., {Hoffman}, R., \& {Sneden}, C. 1993, \pasp, 105, 301

\bibitem[{{Leitherer} {et~al.}(1999){Leitherer}, {Schaerer}, {Goldader},
  {Gonz{\'a}lez Delgado}, {Robert}, {Kune}, {de Mello}, {Devost}, \&
  {Heckman}}]{Leitherer99}
{Leitherer}, C. {et~al.} 1999, \apjs, 123, 3

\bibitem[{{Mackey} \& {Broby Nielsen}(2007)}]{Mackey07}
{Mackey}, A.~D. \& {Broby Nielsen}, P. 2007, \mnras, 379, 151

\bibitem[{{Maraston}(2005)}]{Maraston05}
{Maraston}, C. 2005, MNRAS, 362, 799

\bibitem[{{Maraston} \& {Str{\"o}mb{\"a}ck}(2011)}]{Maraston11}
{Maraston}, C. \& {Str{\"o}mb{\"a}ck}, G. 2011, \mnras, 418, 2785

\bibitem[{{Mathis} {et~al.}(2006){Mathis}, {Charlot}, \&
  {Brinchmann}}]{Mathis06}
{Mathis}, H., {Charlot}, S., \& {Brinchmann}, J. 2006, \mnras, 365, 385

\bibitem[{{McGaugh}(1991)}]{McGaugh91}
{McGaugh}, S.~S. 1991, \apj, 380, 140

\bibitem[{{McMillan} {et~al.}(2007){McMillan}, {Vesperini}, \& {Portegies
  Zwart}}]{McMillan07}
{McMillan}, S.~L.~W., {Vesperini}, E., \& {Portegies Zwart}, S.~F. 2007, \apjl,
  655, L45

\bibitem[{{Milone} {et~al.}(2009){Milone}, {Bedin}, {Piotto}, \&
  {Anderson}}]{Milone09}
{Milone}, A.~P., {Bedin}, L.~R., {Piotto}, G., \& {Anderson}, J. 2009, \aap,
  497, 755

\bibitem[{{Minniti} {et~al.}(2004){Minniti}, {Rejkuba}, {Funes}, \&
  {Kennicutt}}]{Minniti04}
{Minniti}, D., {Rejkuba}, M., {Funes}, J.~G., \& {Kennicutt}, Jr., R.~C. 2004,
  \apj, 612, 215

\bibitem[{{Moeckel} \& {Bonnell}(2009)}]{Moeckel09}
{Moeckel}, N. \& {Bonnell}, I.~A. 2009, \mnras, 400, 657

\bibitem[{{Mucciarelli} {et~al.}(2012){Mucciarelli}, {Origlia}, {Ferraro},
  {Bellazzini}, \& {Lanzoni}}]{Mucciarelli12}
{Mucciarelli}, A., {Origlia}, L., {Ferraro}, F.~R., {Bellazzini}, M., \&
  {Lanzoni}, B. 2012, \apjl, 746, L19

\bibitem[{{Osterbrock}(1989)}]{Osterbrock89}
{Osterbrock}, D.~E. 1989, {Astrophysics of gaseous nebulae and active galactic
  nuclei}

\bibitem[{{Panter} {et~al.}(2003){Panter}, {Heavens}, \& {Jimenez}}]{Panter03}
{Panter}, B., {Heavens}, A.~F., \& {Jimenez}, R. 2003, \mnras, 343, 1145

\bibitem[{{Peacock} {et~al.}(2012){Peacock}, {Zepf}, \&
  {Maccarone}}]{Peacock12a}
{Peacock}, M.~B., {Zepf}, S.~E., \& {Maccarone}, T.~J. 2012, \apj, 752, 90

\bibitem[{{Piotto} {et~al.}(2012){Piotto}, {Milone}, {Anderson}, {Bedin},
  {Bellini}, {Cassisi}, {Marino}, {Aparicio}, \& {Nascimbeni}}]{Piotto12}
{Piotto}, G. {et~al.} 2012, \apj, 760, 39

\bibitem[{{Renzini}(2008)}]{Renzini08}
{Renzini}, A. 2008, \mnras, 391, 354

\bibitem[{{Rubele} {et~al.}(2013){Rubele}, {Girardi}, {Kozhurina-Platais},
  {Kerber}, {Goudfrooij}, {Bressan}, \& {Marigo}}]{Rubele13}
{Rubele}, S., {Girardi}, L., {Kozhurina-Platais}, V., {Kerber}, L.,
  {Goudfrooij}, P., {Bressan}, A., \& {Marigo}, P. 2013, ArXiv e-prints

\bibitem[{{Rubele} {et~al.}(2010){Rubele}, {Kerber}, \& {Girardi}}]{Rubele10}
{Rubele}, S., {Kerber}, L., \& {Girardi}, L. 2010, \mnras, 403, 1156

\bibitem[{{Schweizer} {et~al.}(2008){Schweizer}, {Burns}, {Madore}, {Mager},
  {Phillips}, {Freedman}, {Boldt}, {Contreras}, {Folatelli}, {Gonz{\'a}lez},
  {Hamuy}, {Krzeminski}, {Morrell}, {Persson}, {Roth}, \&
  {Stritzinger}}]{Schweizer08}
{Schweizer}, F. {et~al.} 2008, \aj, 136, 1482

\bibitem[{{Vergely} {et~al.}(2002){Vergely}, {Lan{\c c}on}, \&
  {Mouhcine}}]{Vergely02}
{Vergely}, J.-L., {Lan{\c c}on}, A., \& {Mouhcine}. 2002, \aap, 394, 807

\bibitem[{{Whitmore} {et~al.}(2010){Whitmore}, {Chandar}, {Schweizer},
  {Rothberg}, {Leitherer}, {Rieke}, {Rieke}, {Blair}, {Mengel}, \&
  {Alonso-Herrero}}]{Whitmore10}
{Whitmore}, B.~C. {et~al.} 2010, \aj, 140, 75

\end{thebibliography}
\end{document}